\documentclass[10pt,superscriptaddress,twocolumn,amsmath,amssymb,aps,prl,showpacs]{revtex4-1}
\usepackage{mathrsfs}
\usepackage{array}
\usepackage{graphicx}
\usepackage{dcolumn}
\usepackage{bm}
\usepackage[title,titletoc,header]{appendix}
\usepackage{amssymb}
\usepackage{amsmath}
\usepackage{mathrsfs}
\usepackage{amsmath}
\usepackage{subfigure}
\usepackage{float}
\usepackage[title,titletoc,header]{appendix}
\usepackage{graphicx}
\usepackage{color}

\newcommand{\be}{\begin{equation}}
\newcommand{\ee}{\end{equation}}
\newcommand{\bea}{\begin{eqnarray}}
\newcommand{\eea}{\end{eqnarray}}

\begin{document}
\title{Diffused Vorticity and Moment of Inertia of a Spin-Orbit Coupled Bose-Einstein Condensate}

\author{Sandro Stringari}
\affiliation{INO-CNR BEC Center and Department of Physics, University of Trento, Via Sommarive 14, Povo, Italy}

\date{\today}
\begin{abstract}
 By developing the hydrodynamic theory of spinor superfluids we calculate the moment of inertia of a harmonically trapped Bose-Einstein condensate with spin-orbit coupling. We show that the velocity field associated with the rotation of the fluid exhibits  diffused vorticity,   in contrast to the irrotational behavior characterizing a superfluid. Both Raman-induced and Rashba spin-orbit couplings are considered. In the first case the moment of inertia   takes  the rigid value at the transition between the plane wave and the single minimum phase, while in the latter case the rigid value is achieved in the limit of isotropic Rashba coupling. A procedure to generate the rigid rotation of the fluid and to measure the moment of inertia is proposed. The quenching of the quantum of circulation $h/m$, caused by Raman induced spin-orbit coupling  in a toroidal geometry, is also discussed.

\end{abstract}

\maketitle

{\em Introduction.}
Irrotationality is one of the most important features exhibited by  superfluids \cite{Leggett2006,LS2003}. It is at the origin of phenomena of fundamental  relevance that have been confirmed  experimentally both in superfluid helium and in ultracold atomic  gases, like the quenching of the moment of inertia and the occurrence of quantized vortices. At small angular velocities the formation of quantized vortices is energetically inhibited and the moment of inertia of a superfluid, in the presence of isotropic confinement,  vanishes  at zero temperature as a consequence of the irrotationality of the velocity field. In Bose-Einstein condensates (BECs) the condition of irrotationality is usually associated with the phase  $\phi$ of the order parameter, whose gradient fixes the superfluid velocity according to the    relationship ${\bf v} = (\hbar/m)\nabla \phi$, where $m$ is the mass of atoms.  However, even at $T=0$,  the above irrotationality condition is violated in the presence of spin-orbit coupling, as we will discuss in the present Letter, with important consequences on the rotational behavior of the system.  

 Raman-induced spin-orbit coupled Bose gases, characterized by equal Rashba and Dresselhaus coupling, have been  the object of systematic experimental and theoretical work in the recent years, following the pioneering experimental realization of  \cite{Lin2009}  (for recent reviews see, for example, \cite{YunLi2015,Zhai2015}).  These configurations are characterized by the breaking of Galilean invariance which is responsible for  important consequences like, for example, the breakdown of   Landau's criterion for the critical velocity  \cite{Zhu2012,Zheng2013,Ozawa2013}. 
 Using Bogoliubov theory and sum rule arguments it has been recently shown \cite{HKTN} that the normal (nonsuperfluid) density  of these systems  does not vanish at zero temperature, as happens in usual superfluids. In particular the normal density was calculated as a function of the Raman coupling produced by two lasers that couple two different  hyperfine states, transferring momentum $\hbar {\bf k}_0$. The effect is  largest at the transition between the plane-wave phase and the single-minimum  phase, where the motion of the fluid along the axis  fixed by  ${\bf k}_0$,   completely loses its superfluid nature even at $T=0$, despite the fact that the system is practically fully  Bose-Einstein condensed \cite{Zheng2013}.

The purpose of this letter is to  investigate the rotational properties of   spin-orbit coupled Bose-Einstein condensed gas employing a generalized hydrodynamic formalism, allowing for analytic solutions in the presence of harmonic trapping.   In particular   we  prove the existence of rotational solutions, characterized by diffused vorticity  and we calculate   the corresponding value of the moment of inertia \cite{Fetter}. Both Raman-induced and Rashba spin-orbit coupled Bose gases will be considered. 

{\em Hydrodynamic formalism.}
Our starting point is based on a variational formulation of the Gross-Pitaevskii equations   in terms of the densities  and the phases of the two components of the spinor wave function
\begin{equation}
\label{ansatz}
|\psi\rangle=\left( \begin{array}{c} \sqrt{n_1(\textbf{r})}e^{i\phi_1({\bf r})} \\ \sqrt{n_2(\textbf{r})}e^{i\phi_2({\bf r})} \end{array} \right) \; .
\end{equation}
We will first consider the following Raman-induced 1D spin-orbit single-particle Hamiltonian  
\begin{equation}
H_{sp}= \frac{1}{2m}\left(p_x-\hbar k_0\sigma_z \right)^2 + \frac{1}{2m} p^2_y + \frac{1}{2m} p^2_z - \frac{\hbar\Omega}{2}\sigma_x \;,
\label{Hsp}
\end{equation}
where $\bf{p}= -i\hbar \nabla$ is the canonical momentum and $V_{ext}$ is the external potential. 
The two-body interaction can be written in terms of the intraspecies and interspecies coupling constant as
 $V_{\rm int}=1/2\sum_{\alpha\beta}\int d{\bf r} g_{\alpha\beta}n_\alpha({\bf r})n_\beta({\bf r})$, where $g_{\alpha\beta}=4\pi a_{\alpha\beta}$ are the various coupling constants in different spin channels, with $a_{\alpha\beta}$  the corresponding scattering lengths ($\alpha, \beta=1,2$ label the relevant internal hyperfine Zeeman states). In the following we will assume that   the coupling constants are equal ($g_{\alpha\beta}\equiv g$). In most of alkali atoms this is a good approximation \cite{YunLi2012} for the description of the quantum phases here considered.   
In the single-particle Hamiltonian (\ref{Hsp}) we have omitted  the Raman detuning term $(\delta/2)\sigma_z$ which can be set equal to zero with a proper choice of the frequency shift of the two lasers.  

The Raman-induced spin-orbit Hamiltonian (\ref{Hsp}) is known to predict a second-order phase transition, occurring at the value $\Omega_{cr}= 2\hbar k^2_0/m$ of the Raman coupling, between the so-called plane wave  phase, where all the bosons occupy a state with wave vector   $k_1 =\nabla_x\phi_1=\nabla_x\phi_2=k_0\sqrt{1-(\Omega/\Omega_{cr})^2}$, and the zero-momentum (or single minimum) phase where $k_1=0$. The plane-wave phase reflects the fact that the single-particle excitation spectrum exhibits a double minimum for $\Omega < \Omega_{cr}$ which reduces to a single minimum at   the transition. The second-order transition between the two phases  was   observed experimentally in \cite{Lin2011}.

As shown in \cite{HKTN} the superfluid behavior of the spin-orbit coupled Bose is deeply affected by spin-orbit coupling. This is the consequence of the appearance  of a gap in the excitation spectrum that follows from the breaking of gauge symmetry invariance with respect to the relative phase between the two spin components. The presence of the gap  has the important  consequence of  blocking the relative phase  of the two spin components ($\phi_1=\phi_2\equiv \phi$) not  only in the ground state, but also in the study of the low-frequency  collective oscillations (see Supplemental Material).  This    corresponds to the hydrodynamic regime that will be employed to calculate the moment of inertia. In the hydrodynamic  regime the energy associated with the Raman-induced spin-orbit  Hamiltonian then depends on three variables: the total density $n= n_1+n_2$, the relative density (spin magnetization) $s_z=n_1-n_2$  and the   phase $\phi$ \cite{Martone2012}. Neglecting quantum pressure terms proportional to the density gradients, the energy functional  takes the simplified form $E= \int d{\bf r}\epsilon({\bf r})$ with the energy density given by  
\begin{equation}
\label{E}
 \epsilon({\bf r}) = \frac{\hbar^2}{2m} (\nabla \phi)^2n -\frac{\hbar k_0}{m} (\nabla_x\phi) s_z +    
\frac{1}{2}gn^2 + V_{ext} n -\frac{\hbar\Omega}{2}n
+ \frac{\hbar}{4}\Omega \frac{s_z^2}{n} \; . 
\end{equation}
where, in the calculation of  the Raman term $-\hbar \Omega \sqrt{n_1n_2}\cos(\phi_1-\phi_2)$, we have set $\phi_1=\phi_2 = \phi$ and included only the term quadratic in $s_z$. We have  furthermore considered the single- minimum phase where, at equilibrium, $\nabla \phi =0$ and $s_z=0$.  The extension of the formalism to the plane-wave phase, where the equilibrium regime corresponds to  nonvanishing values of $\nabla \phi$ and  $s_z$, is straightforward \cite{Martone2012,notepw} and the main results  will be given later.

Although the calculation of the moment of inertia does not require a  time-dependent approach,  it is nevertheless instructive to derive the general time-dependent hydrodynamic equations of motion which are obtained  by applying the variational procedure $\delta A=0$ to the action $A= \int dt\int d{\bf r}\hbar \left[\epsilon({\bf r})+\hbar   n\partial_t\phi\right]$.

  Variation of the action with respect to the phase $\phi$ yields the equation of continuity
\begin{equation}
\label{cont}
\partial_t n +\frac{\hbar}{m}\nabla \cdot \left(n\nabla \phi
\right) -\frac{\hbar k_0}{m}\nabla_x s_z =0
\ ,
\end{equation} 
where one recognizes the spin contribution  that modifies the definition of the current density along the $x$ direction according to 
\begin{equation}
\label{jx}
j_x= \frac{\hbar}{m}n\nabla_x \phi -  \frac{\hbar k_0}{m}s_z
\end{equation} 
and is no longer uniquely fixed by the gradient of the phase of the order parameter,  as happens in usual condensates.  The spin contribution to the current actually can cause  the violation of the irrotational nature of the velocty field ${\bf v}= {\bf j}/n$.

Variation of $A$ with respect to the total density yields  the equation for the phase $\hbar\partial_t \phi  -\hbar \Omega/2+gn +V_{ext}=0$,
where we have ignored terms quadratic in the gradient of the phase and in $s_z$.
Finally,  variation with respect to the spin density yields the novel equation 
\begin{equation}
-\frac{\hbar k_0}{m}\nabla_x\phi +\frac{1}{2} \Omega  \frac {s_z}{n}=0
\label{F2}
\end{equation}
which fixes a nontrivial relationship between  the gradient of the phase and the spin density. By releasing the condition $\phi_1=\phi_2$, this latter equation would contain the time derivative of the relative phase $\phi_1-\phi_2$; for this reason it will be called the equation for the relative phase.

The above  equations form a self-consistent set of equations accounting for the dynamics of a spin-orbit coupled Bose gas.  They can be used to calculate the low frequencies excitations of the system, in both  uniform and trapped configurations as well as  the  response  to slowly varying macroscopic perturbations.  In the limit of small amplitude oscillations  they   reduce to a single equation for the density fluctuations  \cite{Martone2012}
\begin{equation}
\label{HD}
m\partial_t^2 \delta n -\nabla_\perp \cdot \left(gn\nabla_\perp \delta n\right)- (m/m^*)\nabla_x  \left(gn\nabla_x \delta n\right) =0
\end{equation}
which generalizes the hydrodynamic formalism   of \cite{Stringari1996},  widely employed in the literature to describe the collective oscillations of trapped Bose-Einstein condensates in the absence of spin-orbit coupling \cite{LS2003}.  
In this form the equation holds   both in   the single-minimum phase ($\Omega>\Omega_c$),  where  the density-independent effective mass is given by $m/m^*  =1-\Omega_{cr}/\Omega$,  and in the  plane-wave phase ($\Omega<\Omega_c$)   where  $m/m^*=1-(\Omega/\Omega_{cr})^2$ \cite{Zheng2013}. Equation (\ref{HD}) provides a useful description of  the  low-frequency ($\omega \ll \Omega$) collective oscillations of the spin-orbit coupled gas in the presence of harmonic trapping, including the most important dipole oscillation along the  $x$ axis for which one finds the result $\omega_d=\sqrt{m/m^*}\omega_x$ with $\omega_x$ the oscillator frequency, as well as the propagation of sound in uniform matter along the spin-orbit axis, characterized by the dispersion law $\omega= q \sqrt{gn/m^*}$.  

{\em Moment of inertia.}  The hydrodynamic formalism developed above is well suited to discuss the  response to  external fields.  For example    the linear response to a static  transverse current perturbation of the form
$H_{pert} = -\lambda P_xe^{iqy}$, where $P_x=-i \hbar\nabla_x   - \hbar k_0 \sigma_z$
is the $x$-th component of the physical momentum, gives access to  the normal (non superfluid) fraction according to the relationship \cite{Baym1969} $\rho_n/\rho=lim_{q,\lambda \to 0}\langle P_xe^{-iqy}\rangle/\lambda$. 
An explicit calculation, based on the equations of spinor hydrodynamics in uniform matter ($V_{ext=0}$), see Supplemental Material, yields the result 
$\rho_n/\rho = \Omega_{cr}/\Omega$  in the single-minimum phase ($\Omega \ge \Omega_{cr}$) and      $ \rho_n/\rho = (\Omega/\Omega_{cr})^2$ in the plane-wave phase ($\Omega \le \Omega_{cr}$), in agreement  with the       findings of \cite{HKTN}.

The moment of inertia is defined by the static response $\Theta= lim_{\omega_{rot}\to 0} \langle L_z\rangle /\omega_{rot}$ of the system to an angular momentum  constraint of the form $H_{pert}=-\omega_{rot} L_z$
where 
\begin{equation}
L_z= xP_y-yPx= -i\hbar (x\nabla_y-y\nabla_x) +\hbar k_0y\sigma_z
\label{Lz}
\end{equation}
is the angular momentum operator and $\omega_{rot}$ is the angular velocity.  Because of its transverse nature, angular momentum actually shares important analogies with the transverse current coupling considered  above.
The inclusion of the resulting perturbation  $E_{pert}= -\hbar \omega_{rot} \int d{\bf r} \left[n\left({\bf r} \times  {\bf \nabla}\right)_z \phi +k_0ys_z\right]$ in the energy functional (\ref{E}) 
affects the equation of continuity which at equilibrium, where $\partial_t n=0$, takes the form
\begin{equation}
\nabla\cdot \left[\left(\nabla \phi-{\bf \omega}_{rot} \times {\bf r}\right)n\right]-k_0\nabla_x s_z =0 \, ,
\label{contRot} 
\end{equation} 
as well as the equation for the relative phase:  
\begin{equation}
-\frac{\hbar k_0}{m}\nabla_x\phi+\frac{1}{2}\Omega \frac{s_z}{n}-\omega_{rot} k_0y =0  .
\label{F2rot}
\end{equation}
The equation for the total phase, in contrast,  is not affected by the coupling and provides  the usual Thomas-Fermi expression $n=(\mu_0- V_{ext})/g$  for the ground-state density with $\mu_0$ fixed by the normalization condition.

It is now possible to derive analytical solutions of the hydrodynamic equations (\ref{contRot}) and (\ref{F2rot})  in the presence of 3D harmonic trapping:
$V_{ext}= (1/2m) (\omega^2_xx^2+ \omega^2_yy^2+\omega^2_zz^2)$. 
In the absence of spin-orbit coupling the moment of inertia of a superfluid  takes a nonvanishing value at $T=0$ only in the presence of a deformed trap in the plane of rotation and is given by the irrotational value $\Theta _{irrot}=\delta^2 \Theta_{rig}$, where $\delta=\langle (x^2-y^2)\rangle /\langle (x^2+y^2)\rangle$ is the deformation of the atomic cloud and  $\Theta_{rig}= m\int d{\bf r} (x^2+y^2)n$ is the  rigid value of the moment of inertia \cite{inertia96}.
To better exploit the effect of spin-orbit coupling in the following we will assume that the harmonic trapping is isotropic in the $x-y$ plane ($\omega_x=\omega_y$) so that the term proportional to $\omega_{rot}$ in the equation of continuity (\ref{contRot}) identically vanishes. The hydrodynamic equations are easily solved by the ansatz $\phi=\alpha xy$ and $s_z=2\beta y n$. Combining the relation $\alpha= k_0\beta$, provided by the equation of continuity, with the one fixed by  Eq. (\ref{F2rot})  for the relative phase,  one finds the result $\beta= \omega_{rot}k_0/(\Omega - \Omega_{cr}/2)$ giving rise, in the single-momentum phase,  to the rigidlike form
\begin{equation}
\label{v}
{\bf v} = ({\bf \omega}_{rot} \times  {\bf r})\frac{\Omega_{cr}}{2\Omega -\Omega_{cr}} 
\end{equation}
for the velocity field ${\bf v}={\bf j}/n$ and the expression
\begin{equation}
\label{theta}
\frac{\Theta}{\Theta_{rig}}=  \frac{\Omega_{cr}}{2\Omega -\Omega_{cr}} 
\end{equation} 
for the moment of inertia, reflecting the quenching of the superfluid fraction of the system. A similar result can be obtained in the plane-wave phase ($\Omega \le \Omega_{cr}$) where the factor $\Omega_{cr}/(2\Omega-\Omega_{cr})$ of Eqs (\ref{v}) and (\ref{theta}) is replaced by $\Omega^2/(2\Omega_{cr}^2-\Omega^2)$. 
Remarkably, at the transition between the two phases one has ${\bf v} = {\bf \omega}_{rot} \times  {\bf r}$ and $\Theta=\Theta_{rig}$, i.e.  the system rotates like a classical rigid body \cite{differentdependence}.

The rigid rotational flow derived above can be also generated by adding a static spin dipole perturbation in the direction orthogonal to the spin-orbit $x$ axis. This is simply achieved by a spin-dependent shift of the trapping  harmonic potential of the form  $(1/2)m\omega_y^2y^2 \to (1/2)m\omega_y^2(y-y_0\sigma_z)^2$.  
The shift   corresponds to adding a static perturbation of the form 
$H_{pert}=- m\omega_y^2y_0 y\sigma_z$
to the original unperturbed Hamiltonian. Such a y-dependent detuning was actually implemented in \cite{Lin2009}  in order to generate an artificial magnetic field, causing the appearance of quantized vortices in spin-orbit coupled gases.
The crucial point is  that this  perturbation coincides with the most relevant spin   term entering the angular momentum operator (\ref{Lz}).  
One can carry out the variational calculation  by including the perturbative  term $E_{pert}=-m\omega_{ho}^2y_0 \int d{\bf r}ys_z$. Neither the  equation of continuity (\ref{contRot}), nor the equation for the phase (\ref{F2rot}) are affected by the perturbation. The equation for the relative phase is instead modified and can be written in the form (\ref{F2rot}) with the proper identification 
\begin{equation}
\frac{\hbar}{m}\omega_{rot}= \omega^2_y\frac{y_0}{k_0}
\label{Omegaeff}
\end{equation}
which defines   an effective rotational frequency.  One then concludes that
 the spin-dependent shift of the  potential, applied to an isotropically trapped Bose gas in the $x-y$ plane, is exactly equivalent to applying the angular momentum constraint $-\omega_{rot} L_z$ with   $\omega_{rot}$ given by (\ref{Omegaeff}).  
In order to measure the induced  angular momentum, and consequently the moment of inertia,  one can study the precession of the collective oscillations in the x-y plane, caused by the presence of angular momentum \cite{Zambelli1998,LS2003,precession}.
 
{\em Quantized vorticity in a toroidal configuration.}
As a further example of the effects of spin-orbit coupling on the rotational properties of the Bose-Einstein condensed gas we will consider here a toroidal confinement of radius $R$ and assume  that the  spin-orbit coupling is oriented along the azimuthal  direction of the torus \cite{Fetterbis}.  This can be in principle achieved using two Laguerre-Gauss laser beams  coupled to two different hyperfine states and transferring angular momentum $\hbar \ell$ \cite{LG}. One can then  map   the problem into the 1D  spin-orbit configuration discussed above with proper periodic boundary conditions for the phase of the order parameter, ensuring the single-value nature of the wave function \cite{Qu}. The discretization of the single-particle levels can be ignored to the extent that $\ell >> 1$ and the stationary solutions of the hydrodynamic equations   correspond to a uniform 1D density. The only relevant equation  is  then the equation  for the relative phase
which, in the single-minimum phase, takes the form $-\hbar k_0 n \partial_\varphi \phi /mR+\Omega s_z/2=0$ with $k_0=\ell/R$. Looking for solutions  
$\phi= \kappa \varphi$, with $\kappa$ an integer number, the circulation turns out to be quantized  according to the rule
\begin{equation}
\int d{\bf l} \cdot {\bf v}= \kappa \frac{h}{m}\left( 1 -\frac{\Omega_{cr}}{\Omega}\right) \; 
\label{circulation}
\end{equation}
in the single-momentum phase, while   
in the plane-wave phase the factor $\Omega_{cr}/\Omega$  should be replaced by  $(\Omega/\Omega_{cr})^2$.
For large values of $\ell$   the critical value  of the Raman coupling takes the form \cite{Qu1}  $\Omega_{cr} =2\hbar \ell^2/mR^2$.
Result (\ref{circulation})   shows that spin-orbit coupling has the effect of reducing the value of the quantum of circulation and that the reduction is total at the transition ($\Omega=\Omega_{cr}$).  It is worth noticing that interference patterns associated with the presence of a persistent current in the torus, being associated with the quantized behavior of the phase, will exhibit the same dislocation structure \cite{Corman2014} exhibited  in the absence of spin orbit coupling. The observation of the reduction of the quantum of circulation requires the direct measurement of angular momentum carried by the current employing, for example, Doppler  techniques \cite{Kumar2016}.
  
{\em Rashba Hamiltonian.}
The hydrodynamic formalism  is naturally applicable also to the case of the 2D spin-orbit   Rashba Hamiltonian
\begin{equation}
H^{R}_{sp}= \frac{1}{2m}\left(p_x-k_x\sigma_x \right)^2 + \frac{1}{2m}\left(p_y-k_y\sigma_y \right)^2 + \frac{1}{2m} p^2_z +V_{ext}
\label{HRsp}
\end{equation}
which is attracting a great interest because of  the non-Abelian nature of the associated gauge potential \cite{Zhai2015}. In the following we will consider the case $k_y \le k_x$ and we will assume that the ground state corresponds to the plane-wave phase where bosons occupy the same single particle state with canonical momentum $p_x= \hbar k_x$ \cite{meanfield}. Similarly to the case of the Raman-induced spin-orbit Hamiltonian (\ref{Hsp}) the Rashba Hamiltonian also exhibits a gap in the excitation spectrum occurring at $\Delta=  2\hbar^2 k_x^2/m$. In this case the gap has the consequence of blocking to zero the value of the $z$ component of the spin density ($s_z=n_1-n_2$) not only in the ground state, but also in the low-energy excitations  (see Supplemental Material). Employing the same ansatz (\ref{ansatz}) for the order parameter one can derive the expression 
\begin{equation}
\label{ER}
\epsilon^R({\bf r}) =\frac{\hbar^2}{2m} (\nabla \delta \phi)^2n -\frac{\hbar^2k_y}{m} (\nabla_y \delta \phi)\phi_Rn  + \frac{\hbar^2 k^2_x}{2m}\phi_R^2 n + \frac{1}{2}gn^2 + V_{ext} n    
\end{equation}
for the hydrodynamic energy density, where we have ignored unimportant constant terms, set $s_z=0$   and, in the calculation of the spin-orbit terms $\langle p_x \sigma_x\rangle= \hbar n(\nabla_x\phi) cos\phi_R$ and  $\langle p_y \sigma_y\rangle= \hbar n(\nabla_y\phi) sin\phi_R$, we have  kept  only terms quadratic in $\delta \phi=\phi-k_xx$ and $\phi_R$.    The resulting energy functional  has a close analogy with the hydrodynamic expression (\ref{E}) holding in the presence of the Raman-induced spin orbit coupling discussed in the previous sections \cite{analogy}. The Rashba
 energy functional is responsible for a  spin contribution to the $y$ component 
\begin{equation}
\label{jy}
j_y= \frac{\hbar}{m}n\nabla_y \phi -  \frac{\hbar k_0}{m}n \phi_R 
\end{equation}
of the current density which  is responsible for the emergence of rotational components in the velocity field.
 The linearized oscillations are described by the same form as in (\ref{HD}), interchanging  the $x$ and $y$ axis  and with the effective mass given by $m/m^*= (1-k^2_y/k^2_x)$.  The hydrodynamic equation then provides the renormalized value $\sqrt{gn/m^*}$ for the sound velocity along the $y$ direction.    In the limit of isotropic spin-orbit coupling the velocity of sound vanishes and the corresponding dispersion law   exhibits a quadratic dependence \cite{BT}. Analogously the dipole frequency along the $y$ direction in the presence of harmonic trapping takes the value
$\omega = \sqrt{m/m^*}\omega_y$.

In a similar way one can calculate the moment of inertia, by adding the constraint
$-\omega_{rot}L_z= \omega_{rot}[i\hbar (x\nabla_y-y\nabla_x) -k_y\sigma_y]$.
In this case one also finds that, in the presence of isotropic harmonic trapping,  the velocity field takes a rigidlike form and the moment of inertia reduces to the expression:
\begin{equation}
\label{thetaR}
\frac{\Theta}{\Theta_{rig}}=  \frac{k^2_y}{2k^2_x-k^2_y} 
\end{equation} 
which takes the classical rigid value in the limit  $k_x=k_y$.

{\em Conclusions.}
We have shown that spin-orbit coupling has a profound effect on the rotational properties of a Bose-Einstein condensate, causing    the violation of irrotationality and the appearance of  rigid flow. This result is the consequence of the breaking of   the current-phase relationship ${\bf j} = (\hbar/m)n\nabla \phi$,  characterizing usual superfluids, and follows from the violation of Galilean invariance. We have described the dynamic behavior of spinor condensates using a spinor hdrodynamic formalism, applicable to both Raman-induded and Rashba spin-orbit coupled BECs. Spin orbit coupling has been shown to give rise to a finite value of the moment of inertia (which can even take the classical rigid value) in isotropic traps at zero temperature, and to cause the reduction of the quantum of circulation in  toroidal configurations. The hydrodynamic results presented in the present Letter  provide a natural basis to investigate the role of phase and spin fluctuations in the presence of spin-orbit coupling and to better understand the conceptual distinction between superfluidity and Bose-Einstein condensation.  The understanding of the connection between the disappearance of superfluidity  in the hydrodynamic behavior, here predicted   for isotropic Rashba coupling,  and the absence of  the Berezinski-Kosterlitz-Thouless transition, recently investigated in two dimensions \cite{Markus} using the same isotropic spin-orbit configuration, is  also expected to provide further insight  into the problem of the superfluid properties of spin-orbit coupled Bose gases and on the role of dimensionality. 

I would like to thank   Lev Pitaevskii and Shizhong Zhang for stimulating discussions and  collaborations. Useful discussions with Markus Holzmann, Giovanni Martone, Tomoki Ozawa and Chunlei Qu are also acknowledged. This work was supported by the QUIC grant of the Horizon2020 FET program and by Provincia Autonoma di Trento.


\begin{thebibliography}{99}
\bibitem{notepw} In the plane-wave phase the spin-polarized ground state  is the result of a  mechanism of spontaneous symmetry breaking which should be stabilized by a (small) external magnetic field in order to apply the formalism of linear response function theory  developed in the present Letter.
\bibitem{noteomega} The effective angular velocity (\ref{Omegaeff})  can be  conveniently written in units of the oscillator frequency as $\omega_{rot}/\omega_y= (y_0/Y) \sqrt{\mu_0/E_r}$ where we have  introduced the Thomas-Fermi radius $Y$, the chemical potential $\mu=m\omega^2_yY^2/2$ and the recoil energy $E_r=\hbar^2 k^2_0/2m$.
\bibitem{inertia96} Analytic results for the moment of inertia are also available for the ideal Bose gas in the presence of harmonic trapping \cite{Sandro96}.
\bibitem{differentdependence}  The different dependence on $\Omega$ exhibited by $\Theta/\Theta_{rig}$ with respect to $\rho_n/\rho$ reflects the fact that the normal density is actually a tensor and that the effects of spin-orbit coupling show up only in the motion along the $x$-axis
\bibitem{precession} The precession of the quadrupole oscillation has been already used to measure the angular momentum of Bose-Einstein condensed gases containing vortical configurations \cite{prec+vortices} as well as the moment of inertia of a trapped superfluid Fermi gas at finite temperature \cite{Grimm}.
\bibitem{LG}  Angular momentum transfer from LG photons has been already employed to produce permanent currents in toroidal configurations \cite{Ryu,Zoran}.
\bibitem{Qu} The   mapping between the spin-orbit coupling created by LG lasers  and the   1D spin-orbit configuration  considered in the present work was  discussed in \cite{Qu1} in the context of the ring  geometry.
\bibitem{meanfield} The plane-wave phase is actually predicted by mean-field theory to be  energetically favored  in the case  $g_{12} <g_{11}=g_{22}$ \cite{ZhaiRashba}. In the case of isotropic interactions ($g_{12} =g_{11}=g_{22}$) the plane wave and the stripe phase are instead degenerate.
\bibitem{analogy} The formal analogy is recovered replacing, in Eqn (\ref{ER}) the $z$ component $s_z$ of the spin density the $y$th component $s_y\sim n\phi_R$.
\end{thebibliography}
\end{document}